\documentclass[twocolumn,  tighten,times ]{aastex63}

\received{2024, March 4}
\revised{2024 May 26}
\accepted{2024 June 14 }
\submitjournal{The Astrophysical Journal}

\shorttitle{UV properties of TTS}
\shortauthors{Nayak et al.}

\graphicspath{{./}{figures/}}

\begin{document}

\title{Simultaneous FUV and NUV observations of T Tauri stars with UVIT/AstroSat: probing accretion process in young stars}

\correspondingauthor{Prasanta K. Nayak}
\email{nayakphy@gmail.com, pnayak@astro.puc.cl}

\author[0000-0002-4638-1035]{Prasanta K. Nayak}
\affiliation{Department of Astronomy and Astrophysics, Tata Institute of Fundamental Research, Mumbai, 400005, India}
\affiliation{Instituto de Astrofísica, Pontificia Universidad Católica de Chile, Av. Vicuña MacKenna 4860, 7820436, Santiago, Chile}
\author[0000-0002-0554-1151]{Mayank Narang} 
\affil{Department of Astronomy and Astrophysics, Tata Institute of Fundamental Research, Mumbai, 400005, India}\affil{Academia Sinica Institute of Astronomy \& Astrophysics, 11F of Astro-Math Bldg., No.1, Sec. 4, Roosevelt Rd., Taipei 10617, Taiwan, R.O.C.}
\textit{}\author[0000-0002-3530-304X]{P. Manoj}
\affil{Department of Astronomy and Astrophysics, Tata Institute of Fundamental Research, Mumbai, 400005, India}
\author{Uma Gorti}
\affiliation{NASA Ames Research Center, MS 245-3, Moffett Field, CA 94035-1000}
\author[0000-0003-4612-620X]{Annapurni Subramaniam}
\affil{Indian Institute of Astrophysics, Bangalore, 560034, India}
\author{Nayana George}
\affiliation{Cochin University of Science and Technology, India}
\author{Chayan Mondal}
\affiliation{Inter-University Centre for Astronomy and Astrophysics, Pune, Maharashtra 411007, India}

\begin{abstract}

We present results from simultaneous FUV and NUV observations of T-Tauri stars (TTSs) in the Taurus molecular cloud with UVIT/AstroSat. This is the very first UVIT study of TTSs. From the spectral energy distribution of TTSs from FUV to near-IR, we show that classical TTSs (CTTSs) emit significantly higher UV excess compared to weak-line TTSs (WTTSs). The equivalent black-body temperatures corresponding to the UV excess in CTTSs ($>10^4$ K) are also found to be relatively higher than that in WTTSs ($<9250$ K). From the UV excess, we have re-classified two WTTSs (BS Tau, V836 Tau) as CTTSs, which has been supported by the follow-up optical spectroscopic study using the Himalayan Chandra Telescope (HCT), showing strong H$\alpha$ line emission. We find that CTTSs show strong excess emission in both FUV ($>$10$^7$) and NUV ($>$10$^3$) bands, while WTTSs show strong excess only in the FUV ($\lesssim$10$^5$), suggesting that excess emission in NUV can be used as a tool to classify the TTSs. We also find a linear correlation between UV luminosity (a primary indicator of mass accretion) and H$\alpha$ luminosity (a secondary indicator of mass accretion) with a slope of 1.20$\pm$0.22 and intercept of 2.16$\pm$0.70.  
\end{abstract}


\section{Introduction} \label{sec:intro}

T Tauri stars (TTSs) are low-mass pre-main-sequence (PMS) stars, generally categorized into Classical TTSs (CTTSs) and Weak-line TTSs (WTTSs) based on their strength of H$\alpha$ emission \citep{Alcala_1993_WTTS, Duvert_2000_WTTS, Gras-Vel_2005_WTTS}. CTTSs show strong and broad H$\alpha$ emission indicating active ongoing accretion from the circumstellar disk onto the central star, while WTTSs show weak and narrow H$\alpha$ emission, suggesting weak or no accretion \citep{herbig1988, martin1998, white2003, barrado2003}.

Circumstellar disks form during the process of star formation due to the conservation of angular momentum and in their short lifetimes ($\sim$ a few Myrs) aid both star and planet formation. 
Material from the disk around the CTTSs is channelled onto the stars along the strong magnetic field lines and generates accretion shocks on the stellar surface, producing hot spots \citep[and references therein]{Hartmann_2016_review}. 
The kinetic energy of the freely in-falling material is dissipated in these accretion shocks. 
Behind these shocks, the energy is converted into radiation which flows back out through the shocks and into the in-falling material. Most of the escaping radiation peaks in the ultraviolet with estimated black body equivalent temperatures of $\sim$ 10$^4$ K \citep{calvet1998}. Thus, accreting stars are characterized by strong UV excesses \citep{gorti2009}.  
The cool and dusty layers of the disk absorb this radiation and re-emit at the infrared wavelengths. 
Hence, the accretion process causes the release of excess energy not only in the UV but also in the IR regions of the spectral energy distribution of a CTTS.  
The UV emissions mainly originate from the hot spots produced by accretion shocks, while near-IR (NIR) and mid-IR emissions come from smaller disk radii while far-IR comes from the midplane of the outer disk.  
As the disk disperses, the SED gradually shows less excess in the IR region \citep{lada1987}. 
Non-accreting WTTS also shows excess in UV due to chromospheric activity and significantly reduced excess in IR due to the presence of a smaller or depleted disk \citep{ingleby2011, schneider2020}.

FUV photons from the star play an important role in heating disk gas and can drive massive thermal winds that eventually deplete disk material. Recent research suggests that FUV-aided dispersal and heating can have consequences for angular momentum transport through magneto-centrifugal winds (e.g., \cite{bai2016}) and the formation of planetesimals \citep{carrera2017}. FUV photons significantly affect disk chemistry—the differences in the molecule content of low-mass TTSs and intermediate-mass Herbig stars are believed to be due to the stronger UV flux from the latter \citep{pascucci2014}. This could have implications for the composition of planetesimals and planets that eventually form in these disks. Even after the disk is dispersed, the bright FUV flux from WTTS could impact the evolution of (proto-) planets and potentially strip their atmospheres \citep{owen2016}. 
Characterizing FUV emission from TTSs and understanding the time evolution of the spectrum is critical to understanding disk evolution \citep{gorti2009}, and more fundamentally, the accretion process itself.

There have been several UV spectral surveys of TTSs by the IUE \citep{valenti2003} and the STIS instruments on the HST \citep{yang2012}. While these studies have yielded valuable information on the FUV spectrum, most of the measurements of the FUV and NUV excesses are not simultaneous. It is thought that the NUV and the FUV flux originate from different regions of the accretion flow. The NUV Balmer continuum arises from the dense pre-shock region of the accretion column and the FUV most likely arises near the hot spot \citep{Ulrich1976, calvet1998, ingleby2013, Hartmann_2016_review}; this makes contemporaneous observations of both bands necessary for understanding the link between the NUV and the FUV emitting regions and the nature of the accretion process \citep{france2014, 2024arXiv240319478N}. UVIT on AstroSat, with its capability of accurate and simultaneous multi-band UV photometry, is the ideal instrument for probing the FUV-emitting regions around young stars. 
The use of multiple filters, three filters each in the FUV ($\lambda_{eff}$ $=$ 1481 \AA, 1541 \AA\ \& 1608 \AA) and NUV ($\lambda_{eff}$ $=$ 2447 \AA, 2632 \AA\ \& 2792 \AA) bands, will give us three-wavelength points allowing us to better reconstruct the flux distribution across FUV and NUV regions.

In this paper, we present simultaneous multi-band observations in FUV and NUV bands of T-Tauri Stars in the Taurus Molecular Cloud. We have modelled the UV flux with a black-body spectrum and have determined the corresponding temperature and luminosity. For a CTTS, the black-body luminosity will provide a direct measure of accretion luminosity. We investigated if there is any significant difference in the black-body temperatures between CTTSs and WTTSs, i.e., between peak temperature for the emission due to accretion and emission due to chromospheric activity. We have also revisited the classification scheme based on the amount of UV excess emitted by TTSs over photospheric emission and the strength of the H$\alpha$ emission. We also search for correlations between UV luminosity and other commonly used accretion/disk tracers such as H$\alpha$ luminosity and NIR excess emission.

The remaining sections of this paper are arranged as follows. In Sect. 2, we discuss the UVIT data, the follow-up optical spectroscopic observation from ground-based telescopes, and the SED analysis of T-Tauri stars. In Sect. 3, we present our results and discuss them. In the Sect. 4, we summarize our results from this study and their implications.

\section{Data selection and Analysis} \label{sec:style}

\subsection{Photometry: UVIT observations}
Four TTSs fields in the Taurus Molecular Cloud (TMC) were observed with UVIT onboard AstroSat (proposal id: A04-210; PI: Annapurni Subramaniam). Each of these fields has a diameter of $\sim$ 28\arcmin~and was centered on the following TTS: FM Tau, V836 Tau, BS Tau, and HD283782. All the observations were carried out using three FUV and three NUV filters in January 2018. The names of the FUV and NUV filters and corresponding exposure times for all targets are listed in \autoref{uvit_obs}. 
FM Tau is observed in the narrow band filter, N279N, centered at the Mg-II line, while other TTSs are observed in the N242W wide band, keeping the other two medium bands (N245M and N263M) as common filters for all the targets.   

The observations were completed in multiple orbits. We applied corrections for spacecraft drift, flat-field, and distortion using the software CCDLAB \citep{postma2017} and created images for each orbit. Then, the orbit-wise images were co-aligned and combined to generate final science-ready images in each filter. Astrometry was also performed using CCDLAB by comparing the Gaia-DR2 source catalog. 
The science-ready images were created for an area of 4K$\times$4K in size with a scale of 0.$''$416/pixel. The details about the telescope and the instruments are available in \cite{subra2016, tandon2017b} and the instrument calibration can be found in \cite{tandon2017a,tandon2020}.

\begin{figure*}[ht!]
\gridline{\fig{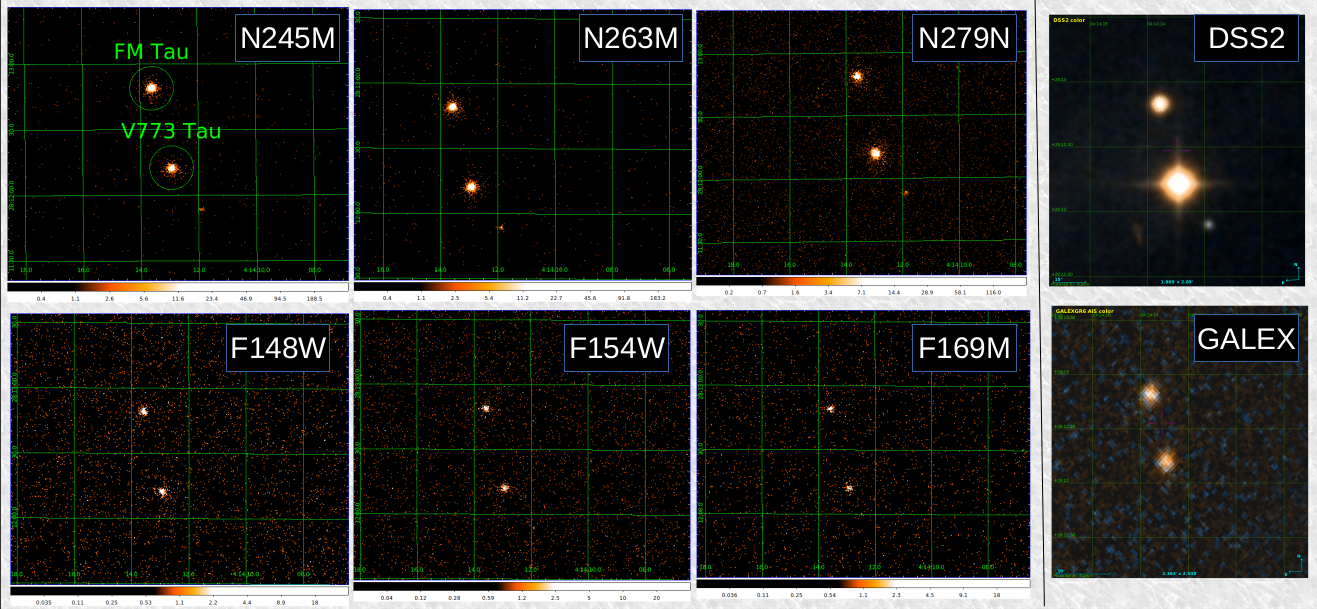}{\textwidth}{}}
\caption{Multi-bands FUV and NUV images of FM Tau and V773 Tau are presented here. The names of the different filters are marked here. DSS2 and GALEX images are also presented for comparison. 
\label{uvit_image}}
\end{figure*}

We used the DAOPHOT tasks and packages in the Image Reduction and Analysis Facility (IRAF){\footnote{IRAF is distributed by the National Optical Astronomy Observatories, which are operated by the Association of Universities for Research in Astronomy, Inc., under a cooperative agreement with the National Science Foundation}} software \citep{stetson1987} to carry out the photometry. 
To detect the sources, we used a threshold of six times the background variation. We performed aperture photometry on the detected stars. We applied saturation corrections to aperture magnitudes and calculated the final magnitudes of the detected stars in the AB magnitude system in the corresponding bands by adding zero-point magnitudes. The values of zero-point magnitudes for the corresponding filters are taken from \citet{tandon2020}. 

Given the large field of view (FOV) of UVIT, it is possible that other TTSs can also be present in the same FOV. To detect these serendipitous TTSs, we cataloged UVIT-detected stars in the various fields and cross-matched them with the Gaia and UV catalogs of the TMC members from \cite{esplin2019}, \citet{Nayak_2023} and \citet{2023arXiv231016820I}  to search for more UV counterparts to TMC members. With UVIT we were able to detect six more TTSs in the field of FM Tau. These are V773 Tau, CW Tau, FO Tau, CIDA 1, Anon 1, and 2MASS J04141188+2811535. While two of the targets (V773 Tau and CW Tau) were detected in both NUV and FUV, four of the TTSs (FO Tau, CIDA 1, Anon 1, and 2MASS J04141188+2811535; hereafter J0414) were only detected in NUV. No additional TTSs were detected in the FOV of BS Tau, V836 Tau and HD 283782. The final UVIT magnitudes along with the corresponding photometric errors in various filters of all the 10 detected TTSs and their coordinates are listed in \autoref{uvit_obs}.

In \autoref{uvit_image}, we show UVIT FUV and NUV images of FM Tau and V773 Tau in different filters. DSS2 and GALEX images of these three sources are also presented in the rightmost panels of Figure~\ref{uvit_image} for comparison. The images clearly show UVIT has better resolution ($\sim$1."4) than the GALEX. Figure \ref{uvit_image} also shows that V773 Tau is brighter than FM Tau in optical (DSS2); however, they appear of similar brightness in UV and their UV magnitudes listed in Table \ref{uvit} also convey the same, indicating that FM Tau exhibits strong excess emission in UV. We discuss the excess UV emissions from all the TTSs in detail in the latter sections.

\subsection{Spectroscopy: HCT observation}

We obtained low-resolution optical spectroscopic of all 10 TTSs. As the optical spectroscopy and UV observations are not simultaneous, it is necessary to obtain multi-epoch spectra of these sources, which will allow us to identify if there is any variability in the H$\alpha$ emission, i.e. variability in the accretion rate. We observed five of our targets with Himalayan Faint Object Spectrograph Camera (HFOSC){\footnote{Further details of the instruments and telescopes are available at http://www.iiap.res.in/iao/hfosc.html}} mounted on the 2~m Himalayan Chandra Telescope (HCT). We also searched the LAMOST database \citep{Cui_2012_LAMOST} and found optical spectra of eight out of eleven of our targets. In \autoref{spectra}, we have listed the names of the sources, from where the spectra are obtained, and observation dates of different epochs for each source. We obtained multi-epoch observations for all of the sources except Anon1 and HD283782.


For the HCT spectra, the wavelength range was covered using Grism 8 (5500–9000\AA) with an effective resolving power of $\sim$1050. Flat frames were taken before each on-target observation, while bias frames were taken both before and after the on-target observation. The FeNe lamp spectra were also obtained after each on-target observation for wavelength calibration. The HCT spectra were reduced in a standard procedure after bias subtraction and flat-field correction using the standard tasks in the IRAF software and the HCT pipeline HAPLI \citep{2023arXiv230812689N}. Then wavelength calibration was performed on extracted spectra for further analysis.

\begin{deluxetable*}{cclccccccc}
\tabletypesize{\scriptsize}
\label{uvit}
\tablecaption{The list of 10 TTS candidates observed with the UVIT.\label{uvit_obs} }
\tablewidth{0pt}
\tablehead{
\colhead{TTS} & \colhead{RA} &  \colhead{Dec}  & \colhead{Filters} & \colhead{exposure} & \colhead{magnitude} &  \colhead{Filters} &  \colhead{exposure} & \colhead{magnitude} & \colhead{Object} \\ 
\colhead{Name} & \colhead{hms} &  \colhead{dms}  & \colhead{(NUV)} &  \colhead{Time (sec)} & & \colhead{(FUV)} &  \colhead{Time (sec)}  & &
\colhead{(Type)}
}
\colnumbers
\startdata
  {} & {} & {}                           & N245M & 1221 & {17.46$\pm$0.02} & {F148W} & {1208} & {19.98$\pm$0.08} & {} \\
FM Tau & 4 14 13.6 & 28 12 49.2  & N263M &  958 & {16.98$\pm$0.02} & {F154W} & {1207} & {19.75$\pm$0.09} & CTTS \\ 
  {} & {} & {}                             & N279N & 1211 & {16.56$\pm$0.04} & {F169M} & {945} & {19.62$\pm$0.11} & {} \\ \hline
  {} & {} & {}                           & N245M & 1221 & {17.80$\pm$0.03} & {F148W} & {1208} & {20.65$\pm$0.11}  & {} \\ 
V773 Tau* & 4 14 13.6 & 28 12 49.2 & N263M & 958 & {16.92$\pm$0.02} & {F154W} & {1207} & {19.94$\pm$0.10} & WTTS \\ 
  {} & {} & {}                             & N279N & 1211 & {15.95$\pm$0.03} & {F169M} & {945} & {20.0$\pm$0.13} & {} \\ \hline 
  {} & {} & {}                           & N245M & 1221 & {16.83$\pm$0.02} & {F148W} &  {1208} & {19.38$\pm$0.06} & {} \\ 
CW Tau* & 04 14 17.0 & 28 10 57.8   & N263M &  958 & {16.24$\pm$0.02}  & {F154W} & {1207} & {19.19$\pm$0.07} & CTTS \\ 
  {} & {} & {}                             &  N279N & 1211 & {15.32$\pm$0.02} & {F169M} & {945} & {18.88$\pm$0.08} & {} \\ \hline 
  {} & {} & {}                            & N245M & 1221 & {19.77$\pm$0.07} & {F148W}  & {1208} & {}  & {}\\ 
FO Tau* & 04 14 49.3 &  28 12 30.46 & N263M  &  958 & {19.05$\pm$0.05} & {F154W} & {1207} & {} & CTTS \\ 
  {} & {} & {}                             &  N279N  & 1211 & {17.95$\pm$0.07} & {F169M} & {945} & {} & {} \\ \hline 
  {} & {} & {}                            &  N245M  & 1221 & {21.58$\pm$0.15} & {F148W} & {1188} & {} & {} \\
CIDA 1* & 04 14 17.520 &  28 06 9.0  & N263M &  958 & {20.61$\pm$0.13} & {F154W} & {1207} & {} & CTTS \\ 
  {} & {} & {}                            &  N279N  & 1211 & {20.08$\pm$0.17} & {F169M} & {945} & {} & {} \\ \hline 
  {} & {} & {}                            &  N245M  & 1221 & {21.48$\pm$0.15} & {F148W} & {1188} & {} & {} \\
J041411.88 & 04 14 11.88 &  28 11 53.31  & N263M &  958 & {20.60$\pm$0.13} & {F154W} & {1207} & {} & CTTS \\ 
{+281153.3*} & {} & {}                     &  N279N  & 1211 & {19.63$\pm$0.14} & {F169M} & {945} & {} & {}\\ \hline 
  {} & {} & {}                            &  N245M  & 1221 & {21.92$\pm$0.18} & {F148W} & {1188} & {} & {} \\
Anon 1* & 04 13 27.216 & 28 16 22.8  & N263M &  958 & {20.82$\pm$0.14} & {F154W} & {1207} & {} & WTTS  \\ 
  {} & {} & {}                            &  N279N  & 1211 & {19.63$\pm$0.13} & {F169M} & {945} & {}\\ \hline 
  {} & {} & {}                           &  N242W & 1214 & {17.92$\pm$0.01} & {F148W} & {1188} & {19.44$\pm$0.06}  & {} \\
BS Tau & 4 58 51.4 & 28 31 24.2  & N245M & 1228 & {18.11$\pm$0.03} & {F154W} & {1190} & {19.24$\pm$0.06} & CTTS \\ 
  {} & {} & {}                            &  N263M & 1225 & {17.83$\pm$0.03} & {F169M} & {1189} & {19.25$\pm$0.07} & {} \\ \hline
  {} & {} & {}                            & N242W  & 1228 & {19.54$\pm$0.03} & {F148W}  & {1199} & {20.68$\pm$0.1} & {} \\
V836 Tau & 5 3 6.6 & 25 23 19.7   & N245M & 1227 & {19.68$\pm$0.06} & {F154W} & {1200} & {20.32$\pm$0.1} & CTTS \\ 
  {} & {} & {}                            &  N263M & 1205 & {19.05$\pm$0.05} & {F169M} & {1187} & {20.18$\pm$0.11} & {} \\ \hline
  {} & {} & {}                             & N242W  & 1209 & {16.20$\pm$0.01} & {F148W} & {1193} & {19.08$\pm$0.05} & {} \\
HD283782 & 4 44 54.4 & 27 17 45.2   & N245M & 1228 & {15.98$\pm$0.01} & {F154W} & {1207} & {18.94$\pm$0.05} & WTTS \\ 
  {} & {} & {}                             & N263M  & 860 & {15.05$\pm$0.01} & {F169M} & {824} & {18.91$\pm$0.08} & {} \\
\enddata
\tablecomments{*In the UVIT field of FM Tau, we found six more TTS candidates having similar exposure times.} 
\end{deluxetable*}

\begin{deluxetable*}{lccccccc}
\tabletypesize{\scriptsize}
\tablewidth{0pt} 
\tablecaption{Spectroscopic observations \label{spectra}}
\tablehead{
\colhead{Source} & \colhead{instruments}& \colhead{observation date} & \colhead{EW(H$\alpha$)} & \colhead{$\overline{EW(H\alpha)}|^{min}_{max}$} & \colhead{L$_{acc}$} & \colhead{$\dot {M}_{acc}$} & Classification\\
\colhead{} & \colhead{} & \colhead{($\rm yyyy-mm-dd$)}& \colhead{} & \colhead{} & \colhead{L$_\odot$} & \colhead{($\times$10$^{-9}$  M$_\odot$yr$^{-1}$)} & \
} 
\colnumbers
\startdata 
Anon 1 & LAMOST & 2014-11-05 & -3.9$\pm$0.2 & $-$3.9 & --- &  --- & WTTS\\ \hline 
BS Tau & HCT    & 2021-01-17 & -31.1$\pm$1.2 & $-$33.1$\pm$2.0 & 0.011$\pm$0.001 & 2.17$^{-0.16}_{+0.16}$ & CTTS\\
       &        & 2021-09-06 & -35.1$\pm$1.3 &  &  &  & \\ \hline
CIDA 1 & LAMOST & 2012-01-04 & -141.9$\pm$34.3 & $-$205.5$^{+107.8}_{-67.7}$ & 0.004$^{-0.002}_{+0.001}$ & 0.70$^{-0.41}_{+0.28}$ & CTTS\\
       &        & 2011-12-18 & -224.8$\pm$21.6 &  &  &  & \\
       &        & 2012-01-13 & -133.3$\pm$16.4 &  &  &  & \\
       &        & 2012-01-22 & -97.7$\pm$7.3 &  &  &  & \\
       &        & 2013-02-08 & -271.7$\pm$18.9 &  &  &  & \\
       &        & 2014-01-25 & -242.1$\pm$28.2 &  &  &  & \\
       &        & 2015-10-13 & -231.5$\pm$26.7 &  &  &  & \\
       &        & 2015-12-28 & -273.2$\pm$47.6 &  &  &  & \\
       &        & 2016-12-31 & -232.9$\pm$11.9 &  &  &  & \\ \hline 
CW Tau & LAMOST & 2014-01-25 & -87.1$\pm$2.3 & $-$126.8$^{+39.7}_{-105.4}$ & 0.54$^{-0.195}_{+0.57}$ & 107.69$^{-38.8}_{+113.65}$  & CTTS \\
       &        & 2014-01-30 & -232.2$\pm$7.2 &  &  &  & \\
       &        & 2014-11-05 & -60.4$\pm$1.8 &  &  &  & \\ \hline 
FM Tau & LAMOST & 2014-01-30 & -112.8$\pm$3.6 & $-$118.8$^{+18.1}_{-24.1}$ & 0.083$^{-0.015}_{+0.021}$ & 16.75$^{-2.99}_{+4.12}$ & \\
\cline{2-4}
           & HCT    & 2021-01-17 & -100.7$\pm$3.7 &  &  &  & \\
           &        & 2021-09-06 & -142.9$\pm$4.2 &  &  &  & \\\hline 
FO Tau & LAMOST & 2011-12-18 & -136.4$\pm$9.3 & $-$143.5$^{+8.3}_{-29.2}$ & 0.059$^{-0.004}_{+0.014}$ & 11.68$^{-0.80}_{+2.88}$ & CTTS  \\
       &        & 2012-01-04 & -172.7$\pm$13.4 &  &  &  & \\
       &        & 2012-01-13 & -135.9$\pm$9.4 &  &  &  & \\
       &        & 2013-02-08 & -135.2$\pm$9.3 &  &  &  & \\
       &        & 2014-01-25 & -137.4$\pm$7.5 &  &  &  & \\
       &        & 2014-01-30 & -158.2$\pm$11.2 &  &  &  & \\ \hline 
HD283782   & HCT & 2021-09-06 & -1.9$\pm$0.1 & $-$1.9 & --- & ---  & WTTS \\ \hline  
V773 Tau   & LAMOST & 2014-11-05 & -3.2$\pm$0.1 & $-$2.5$^{+0.6}_{-0.7}$ & --- & ---  & WTTS \\ 
\cline{2-4}
          & HCT    & 2021-09-06 & -1.9$\pm$0.1 &  &  &  & \\ \hline
V836 Tau   & HCT    & 2021-01-17 & -8.9$\pm$0.5 & $-$13.3$^{+4.4}_{-4.3}$ & 0.013$^{-0.005}_{+0.005}$ & 2.60$^{-1.0}_{+1.0}$ & CTTS \\
           &        & 2021-09-06 & -17.6$\pm$0.5 &  &  &  & \\ \hline
J041411.88 & LAMOST & 2012-01-22 & -319.4$\pm$35.6 & $-$357.4$^{+38.0}_{-47.2}$ & 0.002$^{-0.001}_{+0.001}$ & 0.49$^{-0.01}_{+0.01}$  & CTTS  \\
           &        & 2012-01-23 & -404.6$\pm$138.3 &  &  &  & \\
           &        & 2015-12-28 & -348.2$\pm$69.7 &  &  &  & \\ 
\enddata
\tablecomments{Column 5 represents the average value of ${EW(H\alpha)}$ from multi-epoch observations as listed in column 4 and the uncertainties denote the range of ${EW(H\alpha)}$ from multi-epoch observations. The uncertainties in {L$_{acc}$} and {$\dot {M}_{acc}$} listed in columns 6 and 7, respectively, represent the range in these values due to the range in ${EW(H\alpha)}$ values. }
\end{deluxetable*}

\subsection{Spectral energy distribution} 

Out of 10 UVIT-detected TTSs candidates, only six are detected in both FUV and NUV bands, while the others are detected only in the NUV bands. Therefore, we performed two different approaches to fit the observed SEDs to estimate the stellar parameters. We made a two-component fit to the UV and optical part of the SEDs of the six TTSs detected in both FUV and NUV bands; we used a stellar photospheric model and a blackbody (which represents the excess over the photosphere emission due to accretion shocks) to fit the observed SEDs to determine the excess UV luminosity, its equivalent temperature, and stellar parameters of the stars. We did not include NIR regions for the SED fit as the excess emission in the NIR region comes from the disk. For the remaining four TTSs, we only fit the photospheric models of dwarf stars to the optical part of the observed SEDs to estimate stellar parameters.


To construct observed SEDs, we cross-matched our sample with Gaia-DR3 \citep{Gaia_eDR3_2021, Gaia_DR3_catalog_validation}, APASS DR9 \citep{apass}, and PanSTARRS \citep{magnier_panstarrs_dr2} for the optical; and 2MASS \citep{2mass}
for the NIR region. We also included GALEX GR6/7 data \citep{bianchi2017} for this analysis. 
We have used the virtual observatory (VO) tool, VOSA \citep[VO SED Analyzer;][]{bayo2008} to generate SEDs by converting magnitudes into flux values for corresponding filters and then to fit the observed flux distribution with theoretical model spectra. We have obtained the distances from the Gaia-DR3  \citep{bailer-jones_2021edr3}.  
VOSA performs multiple iterations to get best-fitted spectra to observed flux distribution by varying T$_{eff}$, log(g), metallicity, extinction, and scaling factor ($M_d$) values and gives the best-fitted parameters after performing a ${\chi}^2$ minimization. As these TTSs are in the solar neighbourhood, we consider that they all have similar metallicity as the Sun. We kept extinction as a free parameter to estimate its value from the best-fitted spectra. The scaling factor ($M_d$) is used to scale the model flux to match the observed flux and is defined as $(R_c/D)^2$ where $R_c$ is the radius of the star and D is the distance to the star. For the stellar photosphere, we have used the BT-Settl-CIFIST model \citep{BTSettl2011}.  
We have used log(g) values between 4 to 5 and the full range for T$_{eff}$ from 1200 to 7000 K. In the case of the black body,  we used a range of temperatures from 5000 to 20000 K.

In this study, we determined reduced ${\chi}^2$ for the best fitted SEDs, which is defined as 
\begin{equation}
 {\chi}_{reduced}^2=\frac{1}{N-n}\sum_{k=1}^{N} \frac{(F_{o,k} - M_d \times F_{m,k})^2}{{\sigma_{o,k}}^2}\
\end{equation}
where N is the number of photometric data points, n is the number of input-free parameters, $F_{o,k}$ is the observed flux, and $F_{m,k}$ is the model flux. 
Though $\chi^2$ is used to determine the quality of the fit, often $\chi^2$ values are found to be larger despite visual inspection suggesting them to be well-fitted SEDs. This large $\chi^2$ value arises when the photometric data points have very small observational flux errors (say $<$1\% of observed flux). So, even if the model reproduces the observation apparently well, the deviation can be much higher than the reported observational error \citep[increasing the value of $\chi^2$;][]{Rebassa2021}{}{}. To mitigate this, VOSA has introduced the parameter called the visual goodness of fit (Vgf$_b$) by modifying the $\chi^2$ formula, where the error is considered to be at least 10\% of its observed flux\footnote{http://svo2.cab.inta-csic.es/theory/vosa/helpw4.php?otype=star\&action=help}. The parameter Vgf$_b$ provided by VOSA with a value $\leq$15 is usually considered as a proxy for well-fitted SEDs \citep{Rebassa2021}. We considered the Vgf$_b$ parameter for the goodness of fit, however, we mentioned both $\chi^2$ and Vgf$_b$ parameters from the SEDs fitting process.

\begin{figure*}
\gridline{\fig{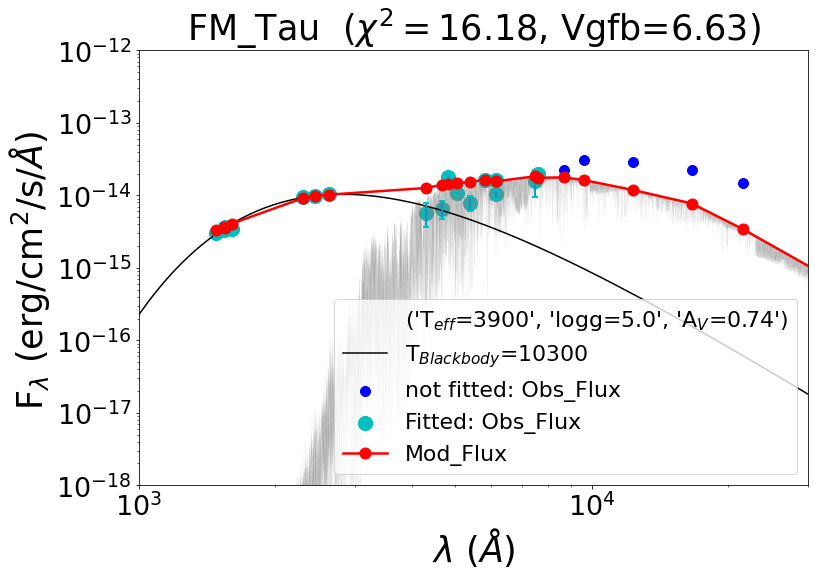}{0.4\textwidth}{(a)}
          \fig{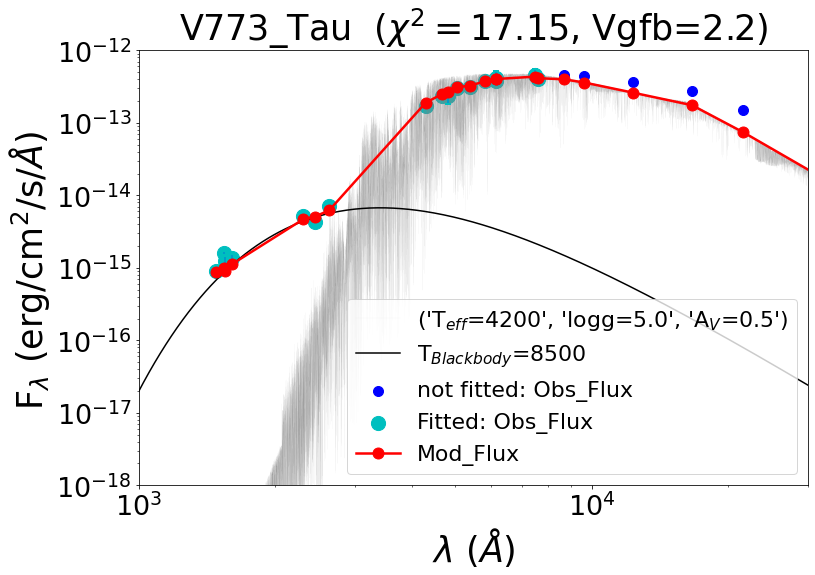}{0.4\textwidth}{(b)}
          }
\gridline{\fig{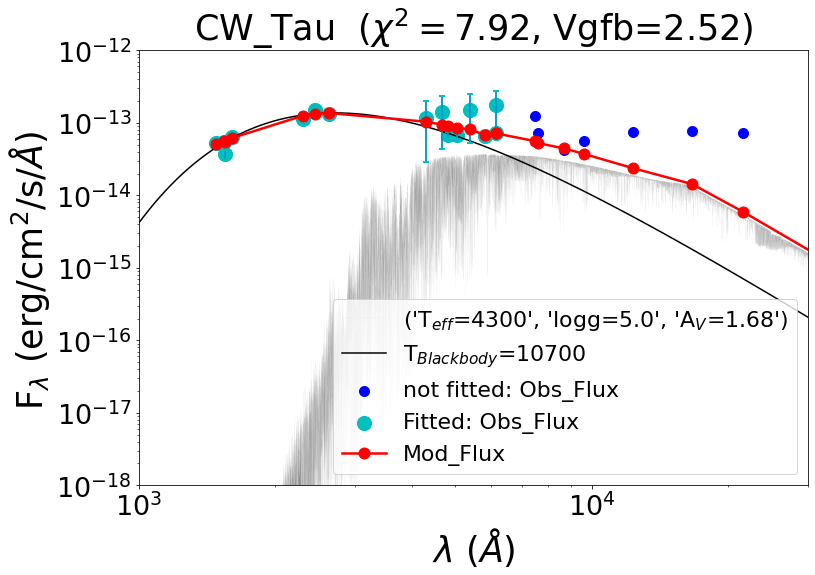}{0.4\textwidth}{(c)}
          \fig{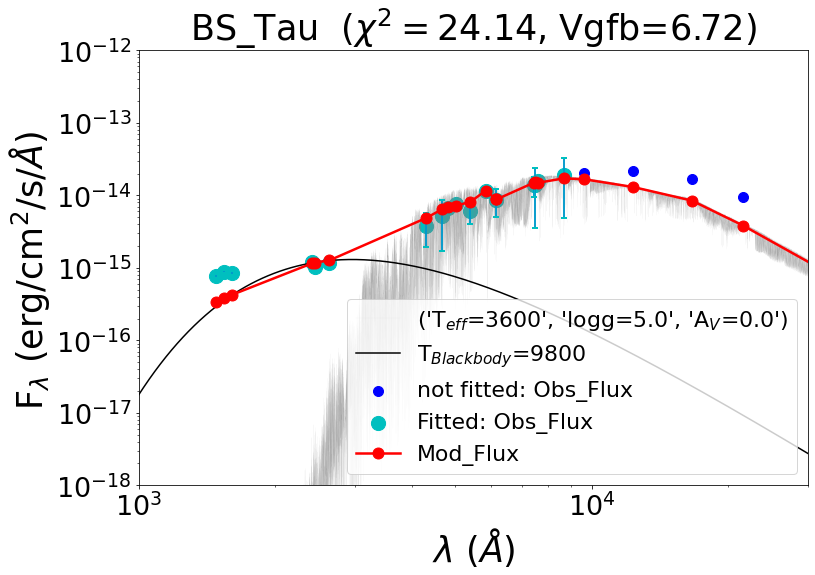}{0.4\textwidth}{(d)}
          }
\gridline{\fig{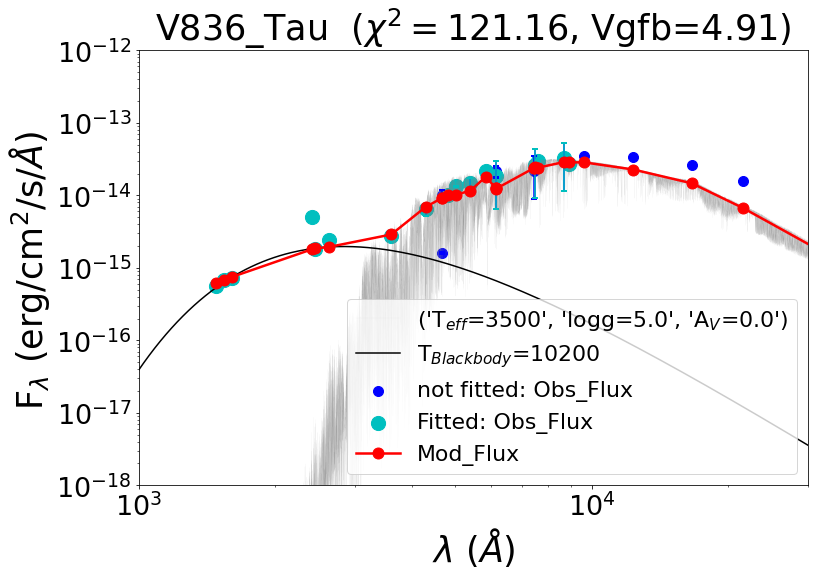}{0.4\textwidth}{(e)}
          \fig{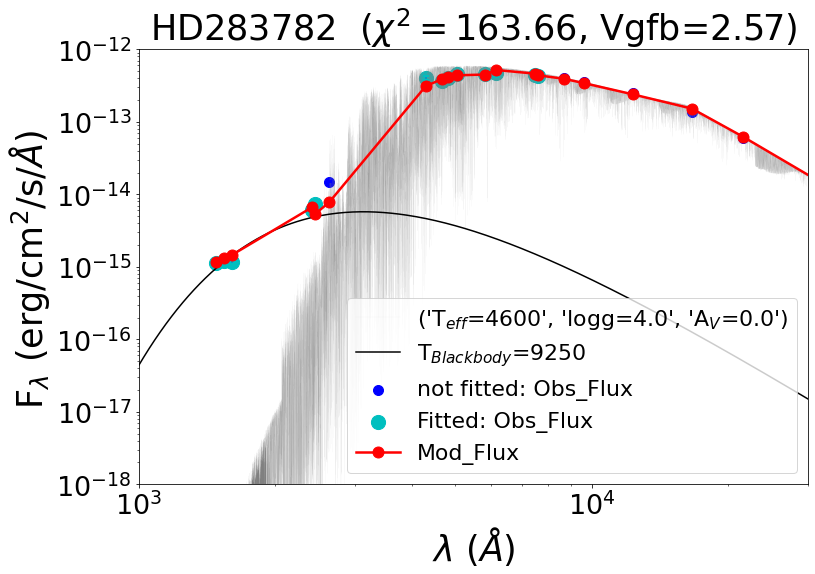}{0.4\textwidth}{(f)}
          }
\caption{SEDs of TTSs are shown here. The names of the TTSs and the best-fitted values of $\chi^2$ and Vgf$_b$ are mentioned on top of each sub-figure. The grey and black lines represent the best-fitted synthetic black body and dwarf spectra, respectively, on observed fluxes (cyan and blue points). Cyan points are included in the fitting algorithm, avoiding the NIR region marked as blue. The red line indicates the expected combined model flux from the fitted synthetic spectra. The best-fitted parameters for dwarf and blackbody spectra are also mentioned in the legend of each sub-figure. 
\label{sed}}
\end{figure*}

In \autoref{sed}, we have shown SEDs of the six TTSs (detected in both FUV and NUV) fitted with model spectra. The observed fluxes are shown as cyan and blue points, where cyan points fall in the UV and optical regions of the energy distribution which is included in the dual component fitting of model spectra. The blue points fall in the NIR region and are not included in the fit. The black and grey lines represent the best-fitted black-body and BT-Settl-CIFIST spectra, respectively, to the cyan points. The combined flux of these two model spectra is shown as the red dashed line. The red points indicate the combined flux in different bands. 
The overlap between red and cyan points indicates the goodness of the two-component fitting. We notice that in the case of HD283782, combined flux matches very well in the NIR region (not included in the fit) as well. Therefore, there is no NIR excess present in HD283782, which indicates that the TTS does not have hot/warm dust around it. The values of reduced $\chi^2$, Vgf$_b$, temperatures, logg, and extinction (A$_V$) corresponding to the best-fitted spectra are noted in the legends. The names of the sources are mentioned at the top of each plot. V836 Tau and HD283782 were found to have high $\chi^2$ values ($>$100) due to the above-mentioned reason of very small photometric error, however, smaller values of Vgf$_b$ ($<$5) indicate that SEDs are fitted well by the model spectra.

In \autoref{sed2}, we have shown SEDs of TTSs with no FUV detections. Due to the absence of FUV points, only the optical region of the observed SED is fitted with synthetic dwarf spectra (BT-Settl-CIFIST) to get the stellar parameters of these TTSs. Fitting two-component spectra to these sources with no FUV data points leads to the wrong estimation of parameters for the blackbody and stellar spectrum. We arrived at this conclusion after examining two-component fits to those six TTSs having both FUV and NUV, but FUV data points were not included in the fitting process and found that the best-fitted spectra were not able to produce the observed FUV flux. 
We have used T$_{eff}$, $log(g)$, and A$_V$ as free parameters, and metallicity to be fixed at Solar value. Blue and cyan points in \autoref{sed2} are observed SEDs, where only cyan points are included in the fitting process. The grey spectrum indicates the best-fitted model spectrum to the optical region of the observed SED. The red point indicated the expected flux due to the best-fitted model spectrum. The best-fitted parameters are mentioned in the legend. Except for the source J0414, all the other three sources have very high $\chi^2$ values, however, all four sources have Vgf$_b$ $\leq$ 15 suggesting the model spectra fitted well to the observed SEDs.



\section{Results and Discussion}

The SED analysis provides us with fundamental parameters of the TTSs, i.e., T$_{eff}$, logg, radius, mass, and bolometric luminosity (L$_{bol}$) of the central stars. As mentioned on the VOSA website, the fitting process and the predicted flux are relatively less sensitive to $\log\ g$. This poses a challenge to put direct constraints on the masses of the components since a slight change in $\log\ g$ can create a large difference in the estimated mass. Therefore, we do not include the estimations of mass and $\log\ g$ in our analysis. The values of other parameters are listed in \autoref{results}.  The luminosity of the black-body provides a direct measure of the accretion luminosity/chromospheric luminosity. In \autoref{results}, we listed the value of bolometric luminosity enclosed within the black-body spectrum (L$_{acc}$) and black-body temperature (T$_{BB}$). We have also estimated the spectral type of each TTS using T$_{eff}$-spectral type relation \citep{pecaut2013} and listed in column 7 of \autoref{results}. We find that TTSs have a T$_{eff}$ range from 3000 K to 4600 K and spectral types of K and M. The T$_{BB}$ ranges from 8500 K to 10700 K. 


\subsection{UV excess as an indicator of accretion}

\begin{figure*}
\gridline{\fig{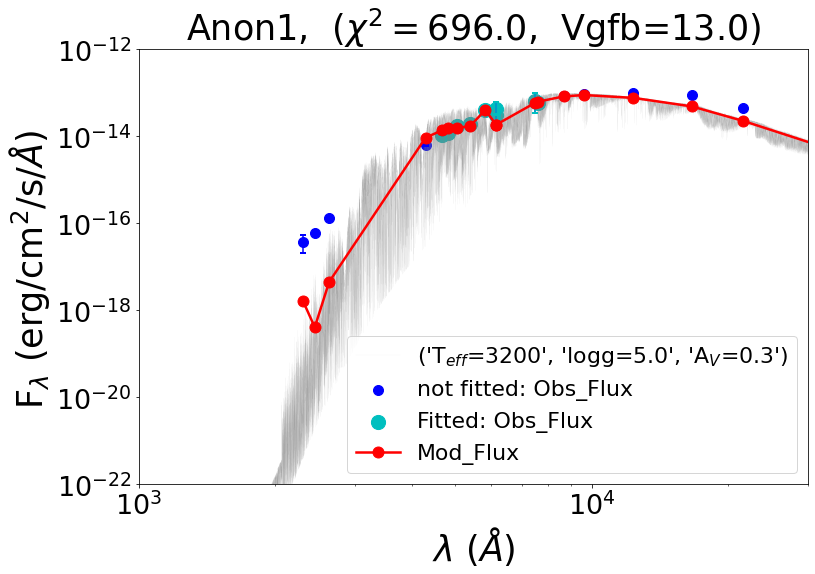}{0.45\textwidth}{(a)}
          \fig{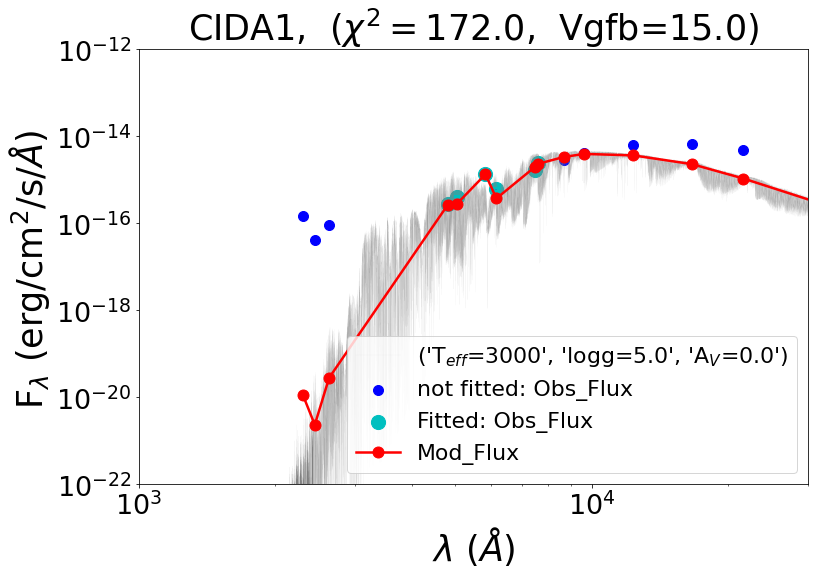}{0.45\textwidth}{(b)}
          }
\gridline{\fig{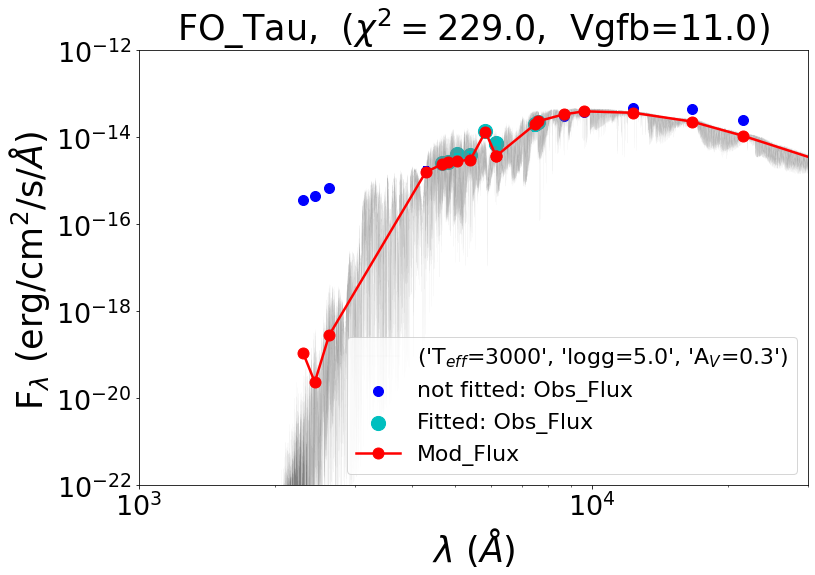}{0.45\textwidth}{(c)}
          \fig{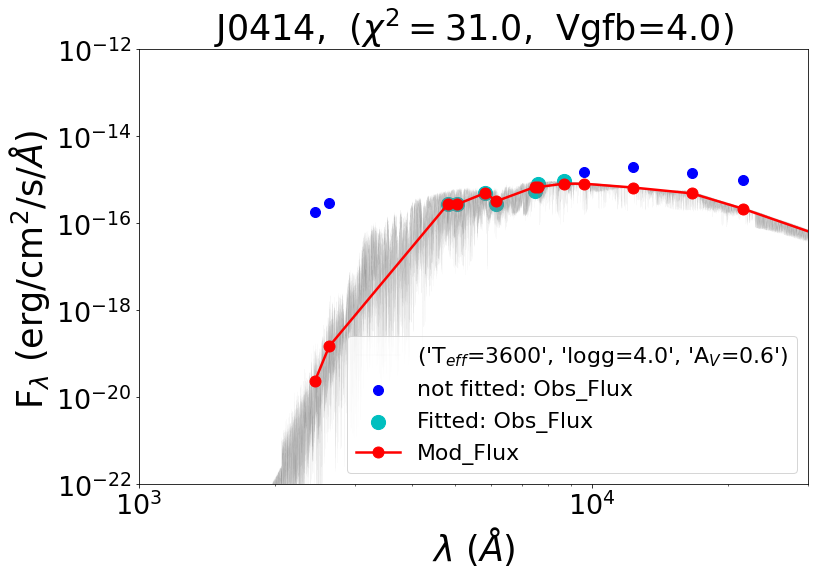}{0.45\textwidth}{(d)}}
\caption{The figure represents the same as Figure \ref{sed} but only the optical region of the SEDs are fitted with theoretical dwarf spectra (BT-Settl-CIFIST), avoiding the UV and NIR regions.
\label{sed2}}
\end{figure*}

We notice in \autoref{sed} that V773 Tau and HD283782 emit negligible amounts of excess emission in NUV bands above the photospheric emission compared to that found in the FUV bands. Whereas, in the other four sources we observe a large excess emission in both NUV and FUV bands. In \autoref{sed2}, we further notice that Anon1 has less excess NUV bands compared to other sources. The presence/absence of NUV excess appears to divide sources into two different categories.

\begin{figure*}[ht!]
\gridline{\fig{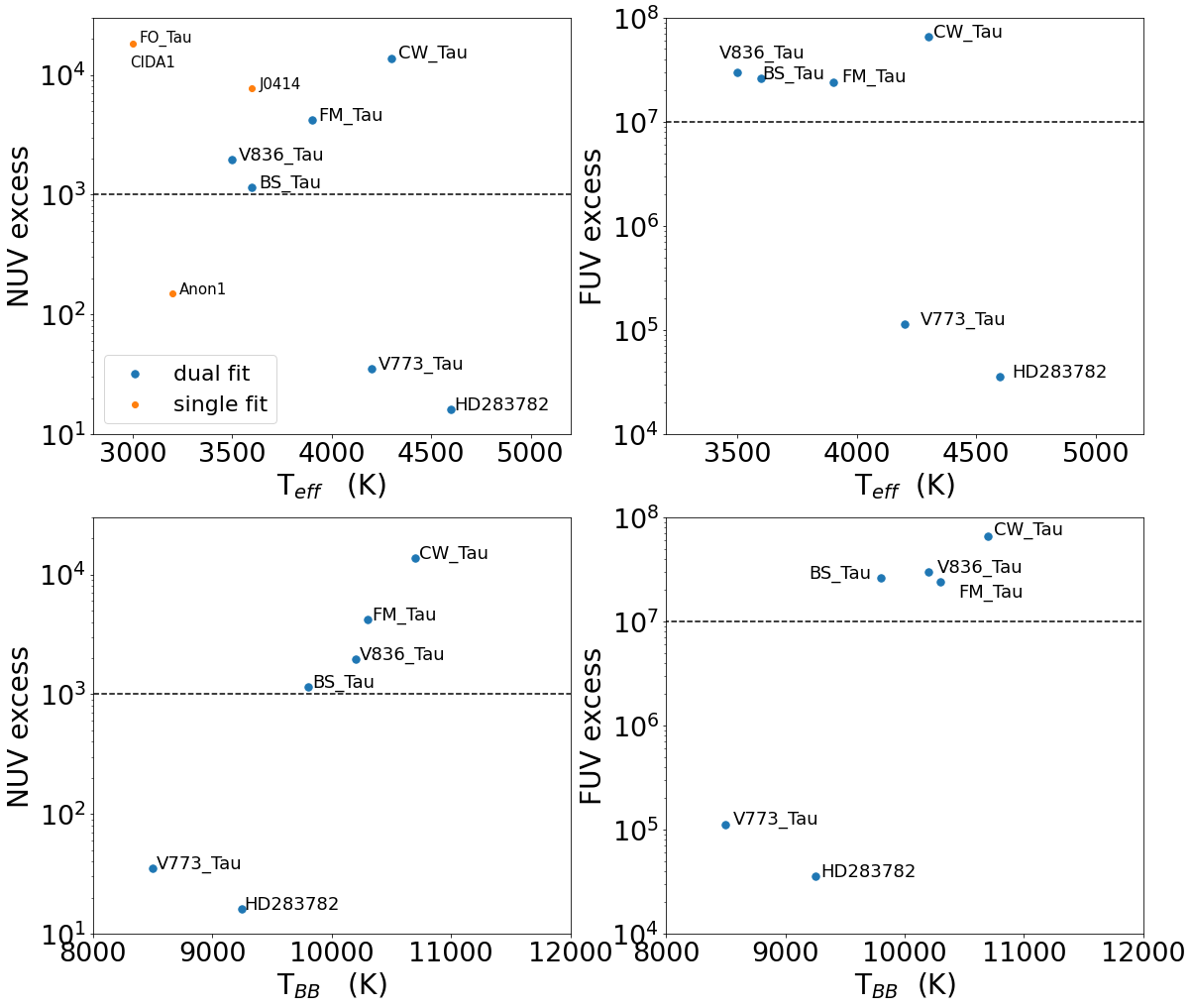}{1.0\textwidth}{}}
\caption{The relations of UV excess as a function of effective temperature (T$_{eff}$; top panel) and blackbody temperature (T$_{BB}$; bottom panel) are shown here. The horizontal dashed line shows the nominal separation between stars with active accretion and chromospheric activity. 
\label{temp_uv_rel}}
\end{figure*}

However, we need to quantitatively estimate the amount of excess emission in all the sources and compare them before giving any conclusion. 
To quantify the excess flux observed in the FUV and NUV regions, we define excess emission in a particular band as  $(\frac{(dereddened~observed~ flux)}{(model ~flux)_{BT-Settl-CIFIST}} - 1)$. To demonstrate UV excess, we used the N245M filter in the NUV and the F148W band in the FUV regions, as these are common bands for all the sources. 
We have plotted NUV and FUV excess as a function of effective temperature (T$_{eff}$) in the upper panel of \autoref{temp_uv_rel} and as a function of blackbody temperature (T$_{BB}$) in the lower panel.
T$_{eff}$ and T$_{BB}$ are estimated from SED fitting. The sources detected in both FUV and NUV are marked as blue points, while the sources with only NUV detection are marked in orange. We can see that there are two groups of population, independent of their spectral types and blackbody temperature, separated by a black dashed line.   
The presence of high UV excess ($>$10$^3$ in NUV and $>$10$^7$ in FUV) suggests that these seven sources (FM Tau, CW Tau, BS Tau, V836 Tau, CIDA1, FO Tau and J0414) could be still actively accreting, while V773 Tau, HD283782 and Anon1 are probable non-accreting WTTSs candidates with comparatively little excess in UV. 
Comparatively low UV excess with almost no NIR excess in HD283782 indicates that it is an example of disk-less WTTS. 
A more careful inspection suggests that stars with higher UV excess (or the probable CTTSs) also tend to have hotter T$_{BB}$ compared to the stars with lesser UV excess. However, we need more sources with simultaneous UV observations to confirm this aspect.  
We classify the stars emitting $>$10$^3$ ($>$10$^7$) excess emission in NUV (FUV) as CTTSs, while the other group of stars are classified as WTTSs.
Our classification based on the UV excess is listed in the \autoref{uvit_obs}. 
However, a strong UV flare might also cause excess UV emission and appear as CTTSs in \autoref{temp_uv_rel}. 
Therefore, we require further evidence from spectroscopic observations before confirming their classification as CTTSs or WTTSs, as CTTSs are expected to show strong H$\alpha$ emission compared to WTTSs.

\begin{deluxetable*}{lccDcccc}[ht!]
\tablecaption{stellar properties estimated from SED analysis\label{results}}
\tablewidth{0pt}
\tablehead{
\colhead{TTS} & \colhead{T$_{BB}$} & \colhead{T$_{eff}$} & 
\multicolumn2c{Radius} &  \colhead{L$_{bol}$} & \colhead{Spectral Type} &  \colhead{L$_{acc}$} & \colhead{$\dot {M}_{acc}$} \\
\colhead{Name} & \colhead{(K)} & \colhead{(K)} & 
\multicolumn2c{(R$_\odot$)} &  \colhead{(L$_\odot$)} & \colhead{} & \colhead{(L$_\odot$)} & \colhead{(10$^{-9}$ $\times$ M$_\odot$yr$^{-1}$)}
}
\decimalcolnumbers
\startdata
FM Tau   & 10300 & 3900 &  {0.73$\pm$0.03}  &   0.11$\pm$0.001 & {K7} & {0.022$\pm$0.0002} & {4.44$\pm$0.03} \\
V773 Tau & 8500 & 4200 &  {3.02$\pm$0.03}  &   2.54$\pm$0.02 & {K6} & 0.018$\pm$0.0002 & ---\\  
BS Tau   & 9800 & 3600 &  {0.96$\pm$0.03}  &  0.14$\pm$0.003 & {M2} & 0.004$\pm$0.0001 & 0.78$\pm$0.03\\  
V836 Tau & 10200 & 3500 &  {1.54$\pm$0.04} &  0.33$\pm$0.003 & {M3} & 0.01$\pm$0.0001 & 1.93$\pm$0.03\\
HD283782 & 9250 & 4600 &  {3.34$\pm$0.05}  &  4.45$\pm$0.028 & {K4}  & 0.073$\pm$0.0006 & ---\\   
CW Tau   & 10700 & 4300 & {0.81$\pm$0.03}  &  0.21$\pm$0.002 & {K5.25} &  0.325$\pm$0.004 & 64.9$\pm$0.09\\ \hline
FO Tau   &      & 3000 & {2.24$\pm$0.07}  &  0.37$\pm$0.001 & {M6.5} &   & \\
Anon 1   &      & 3200 & {2.59$\pm$0.04}  &  0.64$\pm$0.001 & {M5} &   & \\
J041411.88 &      & 3600 &  {0.20$\pm$0.04}  &  0.006$\pm$0.001 & {M2} &  & \\
CIDA 1   &      & 3000 &  {0.61$\pm$0.04}  &  0.027$\pm$0.001 & {M6.5} &  & \\
\enddata
\end{deluxetable*}

\subsection{H$\alpha$ line emission as an accretion indicator}

We analyzed spectra obtained from both the HFOSC/HCT and LAMOST for our sources. We estimated the equivalent width of H$\alpha$ line emission (EW(H$\alpha$)) after subtracting continuum flux from the spectra. In \autoref{spectra}, we have listed the EW(H$\alpha$) along with its measurement error for each source obtained from the multi-epoch observations with different telescopes (column 4). Names of the telescopes and the dates of observations are also mentioned in columns 2 and 3, respectively. 
We notice that EW(H$\alpha$) of FM Tau and V773 Tau from HCT observations matches well with that obtained from the LAMOST spectra. We are unable to obtain a multi-epoch of observation for Anon 1 and HD 283782. 
In column 5, we have listed average values of EW(H$\alpha$) corresponding to each TTS, and errors associated with it indicate the range in EW(H$\alpha$) from multi-epoch observations ( $\overline{EW(H\alpha)}|^{min}_{max}$). All the sources, except J0414, are found to have smaller measurement uncertainties in EW(H$\alpha$) compared to its range from multi-epoch observations. 
As the strength of H$\alpha$ provides an indirect indication of accretion rate, a large range in EW(H$\alpha$) with relatively smaller observational errors in CIDA 1 and CW Tau suggests a large variation in accretion rate in these TTSs.

\begin{figure*} 
\plotone{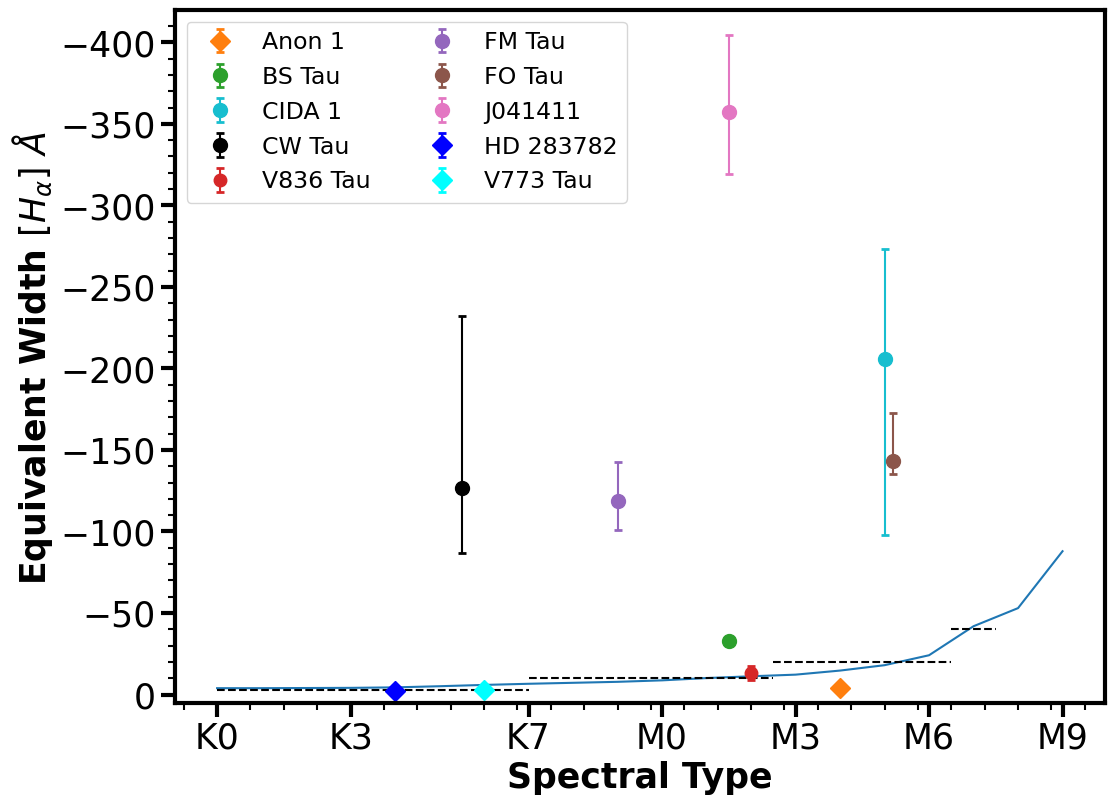}
\caption{The mean H$\alpha$ equivalent widths ($\overline{EW(H\alpha)}|^{min}_{max}$) from multi-epoch observations of TTSs are plotted as a function of their spectral types. The vertical lines represent the range of EW(H$\alpha$) from multi-epoch observations, as listed in \autoref{spectra}. A little offset is applied to FO Tau along the spectral types to avoid its overlap with CIDA 1.  
The solid line \citep{barrado2003} and dashed line \citep{white2003} indicate relations between spectral types and EW(H$\alpha$), which separates CTTSs (circular points) from WTTSs (diamonds). 
\label{halpha_rel}}
\end{figure*}

The H$\alpha$ line strength has often been used to distinguish between CTTSs and WTTSs in the literature \citep{white2003}. WTTSs show weak emission due to chromospheric activity, while comparatively strong emission from CTTSs is produced due to accretion. However, there is no unique value of EW(H$\alpha$) that acts as a dividing line for this classification \citep{barrado2003,white2003}. \citet{barrado2003} provided EW(H$\alpha$) values as a function of spectral types, derived from the observed saturation limit for the chromospheric activity (blue line in Figure \ref{halpha_rel}). Whereas, \citep{white2003} provided maximum EW(H$\alpha$) values for WTTSs for different ranges in spectral types (black dashed line in \autoref{halpha_rel}).
We have over-plotted  $\overline{EW(H\alpha)}|^{min}_{max}$ values of our sample TTSs in \autoref{halpha_rel}.  
The spectral types are estimated from photospheric temperature based on the correlation given in \cite{pecaut2013}, listed in \autoref{results}.  
\autoref{halpha_rel} shows that FM Tau, BS Tau, CW Tau, CIDA 1, FO Tau and J0414 have high values of EW(H$\alpha$) with respect to the expected values from WTTSs with similar spectral types. These stars also show high UV excess, as discussed in the previous section, indicating that their classification as CTTSs is robust. Except for BS Tau, all other sources discussed above show significant variation in the EW(H$\alpha$), suggesting a large variation in the accretion rate. J0414 has a large uncertainty in the estimation of EW(H$\alpha$), so variation in J0414 is most likely not due to accretion variability. In the next section, we estimate and discuss the accretion luminosity and mass accretion rate of these sources. We, however, need more follow-up observations for a detailed study of their variability.

Anon 1, V773 Tau, and HD 283782 have equivalent widths less than that expected for their spectral types, which indicates that these TTSs can be classified as WTTSs, marked as diamond points in \autoref{halpha_rel}. From the SEDs, we have also seen relatively low UV excess over photospheric emission. Both these scenario suggests that these are WTTSs. 
\autoref{halpha_rel} also shows that V836 Tau falls on the line separating CTTSs from WTTSs and the range in the EW(H$\alpha$) extends on both sides, making it difficult to classify. However, we notice a large amount of UV excess emission as shown in \autoref{temp_uv_rel}, which is comparable to that found in other CTTSs, suggesting that V836 Tau is a CTTS.  
We also notice that V836 Tau changes its EW from 8.88 to 17.6 in the two epochs of observation, which corresponds to $\sim$100\% increase in the EW. These observations suggest that the star is going through a variable accretion rate. 
Therefore, based on both UV excess and EW(H$\alpha$), we suggest that V836 Tau is a slowly accreting CTTS with a variable accretion rate. Further monitoring observations of V836 Tau are required to understand its accretion properties. Our classification based on the spectroscopic observations is listed in the \autoref{spectra}.

Our classification of these sources matches well with the previous literature. From the literature, we find that FM Tau, CW Tau and FO Tau are classified as CTTSs, while V773 Tau and HD283782 are classified as WTTSs \citep{ herbig1988, kenyon1998, valenti2003, ingleby2013, McClure_2013, gomez2015}, which agrees with our classification. BS Tau and V836 Tau are classified as WTTSs based on the UV-NIR color-color relation \citep{gomez2015}, contrary to our classification of CTTSs for both sources. Based on HST FUV spectroscopic observations, V836 Tau is classified as a slowly accreting CTTS \citep{ingleby2013}. However, both UV excess and EW(H$\alpha$) values from our analysis indicate that BS Tau and V836 Tau are CTTSs. We did not find any classification for CIDA~1, J0414, and Anon~1 in the literature. Thus from our analysis, we reconfirm that V836 Tau is a CTTS and for the first time we report BS Tau, CIDA~1, and J0414 are CTTSs and Anon~1 is a WTTS. 
This study demonstrates the importance of multi-wavelength SED analysis and in particular, the simultaneous FUV and NUV observations to classify the TTSs as classical or weak-line.

%
\subsection{Relation between H$\alpha$ luminosity and accretion luminosity}

In this study, we have determined excess UV flux by fitting the excess emission with the black-body spectrum. The flux enclosed within the fitted black-body spectrum is nothing but the excess UV emission due to accretion in a CTTS and chromospheric activity in a WTTS. Hence, the bolometric luminosity of the black-body spectrum provides the direct measure of accretion luminosity in a CTTS. We used this luminosity value to calculate the mass accretion rates in CTTSs. 
Mass accretion rate (($\dot {M}$)) and accretion luminosity are related by the following equation by \cite{gullbring1998}:   
\begin{equation}
    L_{acc}/L_\odot = (GM_*{\dot {M}}/R_*) \times (1 - R_* / R_{in})
\end{equation}
where M$_*$ and R$_*$ are stellar mass and radius and R$_{in}$ is the disk truncation radius from which the gas falls onto the star. R$_{in}$ is typically assumed to be $\sim$ 5~R$_*$ \citep{gullbring1998}. The error introduced by this assumption on the measured mass accretion rates, considering that R$_{in}$ for a pre-main sequence star can span from 3 to 8 R$_*$, is less than 20\%. Therefore, the above equation can be written as 
\begin{equation}
\label{Macc_rel}
    {\dot {M}}_{acc} = (1 -  R_* / R_{in}) \times (L_{acc}R_* / GM_*)  
    \sim 1.25 (L_{acc}R_* / GM_*)
\end{equation}

We have considered R$_*$/M$_*$ $\sim$ 5 R$_\odot$ /M$_\odot$ 
and used above equation \ref{Macc_rel} to estimate ${\dot {M}}_{acc}$. 
The values are listed in the last column of \autoref{results}. We have only calculated the mass accretion rate for the stars classified as CTTSs. L$_{acc}$ and $\dot {M}_{acc}$ estimated from excess UV flux provides a direct measure of these parameters.

\begin{figure}[ht!]
\gridline{\fig{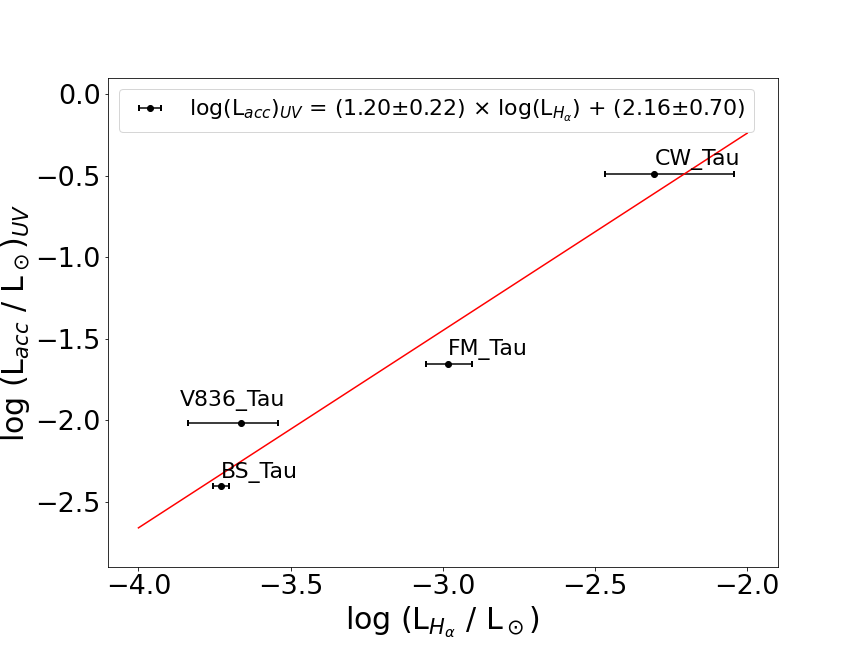}{\columnwidth}{}
}
\caption{The figure shows the relation between accretion luminosity estimated using UV luminosity and ${H_\alpha}$ luminosity $(L_{H_\alpha})$. The uncertainties in $L_{H_\alpha}$ are calculated from the range in EW($H_\alpha$), obtained from multi-epoch spectroscopic observations. The red line represents the linear fit to the data points and the linear relation is noted in the legend. The relatively large residuals are expected since the ${H_\alpha}$ and UV data were not taken simultaneously. 
\label{acc_lum_halpha}}
\end{figure}

EW(H$\alpha$) is often used as a proxy for accretion \citep{white2003}. However, EW(H$\alpha$) is a secondary indicator for accretion. 
Therefore, we tried to find the correlation between L$_{acc}$ from UV excess and $L_{H_\alpha}$. 
We used extinction-corrected Pan-STARRS r band flux density as a proxy for the continuum flux density underlying the H$_\alpha$ line and the distance obtained from Gaia-DR3 to calculate $L_{H_\alpha}$ from average EW(H$_\alpha$) values listed in Table \ref{spectra}. 
The continuum flux density at H$\alpha$ is given as \citep{mathew2018}, 
$$F_{\nu,cont}(H\alpha) = F_{\nu,0} \times 10^{(\frac{-r_0}{2.5})}$$ where $F_{\nu,0} = 3.08 \times 10^{-23}~W\ m^{-2}\ Hz^{-1}$ and r$_0$ is the extinction corrected Pan-STARRS magnitude. The extinction is obtained from the SED fitting as listed in \autoref{results}. 
In \autoref{acc_lum_halpha}, we plotted L$_{acc}$ estimated from UV excess vs that estimated from $(L_{H_\alpha})$, which shows a linear correlation (red line) between them with a slope of 1.20$\pm$0.22 and intercept of 2.16$\pm$0.70. The uncertainties associated to $L_{H_\alpha}$ shown in \autoref{acc_lum_halpha} mainly signifies the range in $L_{H_\alpha}$ values from $\overline{EW(H\alpha)}|^{min}_{max}$.
The uncertainties in L$_{acc}$ or UV luminosities are obtained from the observational errors in flux and in distance estimations, not from the errors in best-fitted model parameters. As both observed flux errors and distance errors are very small, the uncertainties in L$_{acc}$ are also found to be very small. 
There are also previously reported correlations between L$_{H_\alpha}$ and L$_{acc}$ \citep{vogt1994, herczeg2008, dahm2008, fang2009, ingleby2013}. 
We compare our relation with that estimated by previous studies in \autoref{Lacc_rel} and noticed that our relation matches quite well with the slope and intercept values with the previous estimation from the literature.
However, we get a relatively large error in its intercept value, which could be due to the small sample size and non-simultaneous observation of ${H_\alpha}$ and UV. A large sample with simultaneous observations in UV and optical spectroscopy is required for a better correlation. 
We have also used this relation to estimate L$_{acc}$ from L$_{H_\alpha}$ for all the CTTSs listed. This helps us to calculate L$_{acc}$ for the TTSs with no FUV  observations for which we could not estimate UV luminosity. We then also calculated ${\dot {M}}_{acc}$ using \autoref{Macc_rel}. The values of L$_{acc}$ and ${\dot {M}}_{acc}$ are listed in columns 6 and 7 of \autoref{spectra}, respectively. The uncertainties associated with L$_{acc}$ and ${\dot {M}}_{acc}$ indicate the range in these values due to the range in $\overline{EW(H\alpha)}|^{min}_{max}$.
We notice that J0414, CIDA~1, V836 Tau, and BS Tau are the slowly accreting CTTSs with a mass accretion rate a few times 10$^{-9}$ M$_\odot$yr$^{-1}$, while FM Tau, CW Tau, and FO Tau are found to have comparatively higher accretion rate (a few times 10$^{-8}$ M$_\odot$yr$^{-1}$).

 \begin{table}[]
     \centering
     \begin{tabular}{|c|c|c|}
     \hline
         slope & intercept & reference  \\
         (m) & (c) &   \\ \hline
         {1.20$\pm$0.11} & {2.0$\pm$0.4} & \citet{herczeg2008} \\
         {1.18$\pm$0.26} & {2.19$\pm$0.64} & \citet{dahm2008} \\
         {1.25$\pm$0.07} & {2.27$\pm$0.23} & \citet{fang2009}\\ 
         {1.31$\pm$0.03} & {2.63$\pm$0.13} & \citet{manara2012} \\
         {1.0$\pm$0.2} & {1.3$\pm$0.7} & \citet{ingleby2013} \\ 
         {\bf{1.20$\pm$0.22}} & {\bf{2.16$\pm$0.70}} & {\bf this work} \\ \hline
     \end{tabular}
     \caption{Compares the empirical relation between L$_{acc}$ and L$_{H\alpha}$ found in Figure 6 with literature values. The formulation used is $log(L_{acc}/L_\odot) = m\times log(L_{H_\alpha}/L_\odot) + c$.
     }
     \label{Lacc_rel}
 \end{table}

\begin{figure}[ht!]
\gridline{\fig{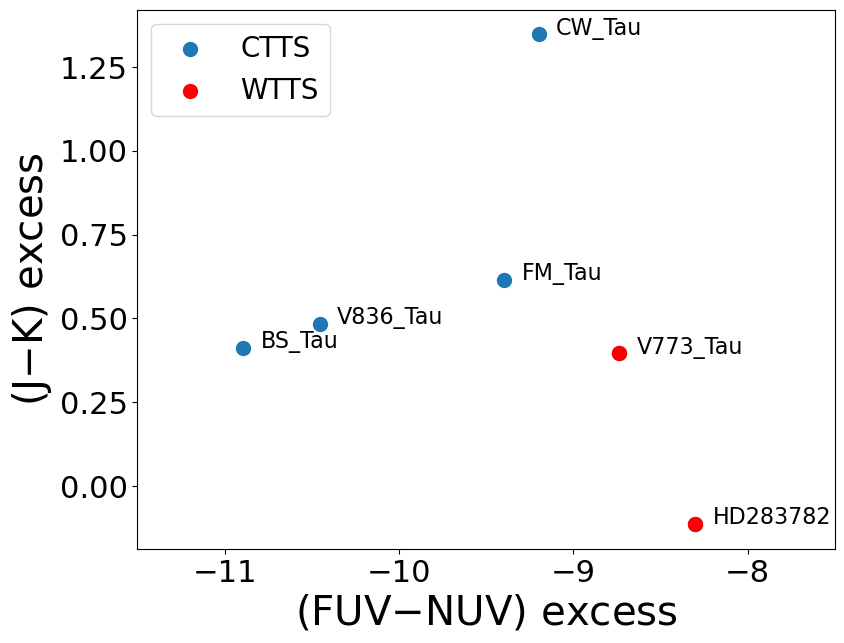}{1.0\columnwidth}{}}
\caption{The figure shows the relation between NIR excess and UV excess. 
\label{ir_uv_rel}}
\end{figure}

\subsection{Relation between UV excess and IR excess}
The NIR color excess is taken as evidence for the presence of circumstellar disks around TTSs. The strength of the infrared color excess indicates the amount of dust and gas in these disks. A strong excess suggests a substantial amount of material in the disk, while a weaker excess may indicate a less massive or more evolved disk. The strong excess in FUV and NUV suggests the presence of significant accretion onto the stars. In contrast, the high FUV and relatively low or no NUV excess suggest that the UV emission originates from chromospheric activity. 
FUV emission from the star plays a significant role in heating gas disks and driving massive thermal winds which deplete the disk material. Hence, excess in (FUV $-$ NUV) color in WTTSs will tell us if the stars are dominant in chromospheric activity, while for CTTSs, it tells us whether the stars are dominant in FUV emission.  
Also, the comparison between UV color and NIR color helps us to identify the changes in disk morphology (primordial gas-rich disks or evolved or transitioning to a debris disk) from FUV bright stars to FUV faint stars.
The UV color index is plotted against J$-$K infrared color excess from 2MASS observations in \autoref{ir_uv_rel}. J$-$K excess is considered to be a good estimator of the accretion rate \citep{meyer1997}. 
The color excess is defined as: 
$-2.5\times[log(\frac{(observed~ flux)_{Band1}} {(observed flux)_{Band2}}) -  log(\frac{(expected~flux)_{Band1}} {(expected~flux)_{Band2}}$)], in magnitude unit. 
For CTTS, we notice that stars with lower NIR color excess,  i.e., with weaker inner disks, 
appear bluer in the (FUV $-$ NUV) color, which means they have stronger FUV emission. 
This suggests that strong FUV emission might have caused the depletion of the gas and dust in the circumstellar disk in CTTSs.
The WTTSs seem to be following a different slope relation compared to CTTSs but with two WTTSs data points, it is not so conclusive. A larger sample of CTTSs and WTTSs is required to better understand this relation.  


\section{Summary}
In this study, we present the accretion properties of 10 TTSs in the TMC, namely FM Tau, BS Tau, V836 Tau, HD283782, V773 Tau, CW Tau, FO Tau, Anon 1, CIDA 1, and 2MASS J04141188+2811535. This is the first UVIT study of T Tauri stars and highlights the significance of simultaneous multi-band UV observations of TTSs to study their accretion properties. Six of 10 are detected in both FUV and NUV bands, while the remaining four are detected only in NUV. We report, for the first time, the  UV photometry of 2MASS J04141188+2811535 (not detected in GALEX, which could be due to short exposure time). 

For the sources detected in both FUV and NUV, we modelled excess UV flux emitted by TTS as black-body emission and measured the excess UV emission by two-component SED fitting to the UV and optical regions of the observed SED. 
From the SED fit, we obtained fundamental stellar parameters (temperature,  extinction, radius, and bolometric luminosity) of these TTSs. We also estimated black-body temperature and luminosity corresponding to the excess UV emission. For the sources with only NUV detection, we only fit the optical region of the SEDs to obtain the stellar parameters and excess emission in NUV bands.

We noticed there are two categories of TTSs based on the excess UV emission: one shows strong excess emission ($>$10$^3$ in NUV and $>$10$^7$ FUV) over the photosphere; the other with excess emission mostly in FUV ($\lesssim$10$^5$) and a little excess in NUV ($\lesssim$10$^2$). We classify the TTSs with strong excess emission in both FUV and NUV as CTTS, and the other group as WTTS. This study shows that UV excess can be used as a tool to distinguish CTTS and WTTS.

We found that the classification of TTS as classical or weak-line based on the spectral type vs EW(H$\alpha$) relation \citep{barrado2003, white2003} matches well with that made based on UV excess.
Our classification also matches well with the literature. From our analysis, we reconfirm that V836 Tau is CTTS. For the first time, we report that BS Tau, CIDA~1 and 2MASS J04141188+2811535 are still actively accreting and classify them as CTTS, and Anon 1 as non-accreting or WTTS.


We found that H$\alpha$ luminosity and accretion luminosity (from UV luminosity) are linearly correlated with a slope of 1.20$\pm$0.22 and an intercept of 2.16$\pm$0.70. The correlation also matches well with previous estimates by \citep{herczeg2008, dahm2008, fang2009, manara2012}. From the mass accretion rate calculation based on UV luminosity and H$\alpha$ luminosity, we found that CW Tau, FM Tau and FO Tau are going through a strong accretion phase with an accretion rate $>10^{-8}$ M$_\odot$yr$^{-1}$, 
while other CTTS are accreting at a slower rate ($\sim$ a few times 10$^{-9}$ M$_\odot$yr$^{-1}$).  
The study brings out the importance of multi-wavelength SED analysis and simultaneous FUV and NUV observation to get a better estimation of the accretion luminosity and accretion rate of TTS.

Comparing the UV color with the NIR color, we notice that stars with higher NIR color excess, i.e., the presence of a substantial amount of material in the disk, appear redder in the (FUV $-$ NUV) color, and vice versa. This suggests that stronger FUV emission might have caused the depletion of the gas in the disk. 
However, a study with a large number sample is required for a better understanding of this relation.

\acknowledgments

This publication uses UVIT data from the AstroSat mission of the ISRO, archived at the Indian Space Science Data Centre (ISSDC). The UVIT project is a result of collaboration between IIA, Bengaluru, IUCAA, Pune, TIFR, Mumbai, several centres of ISRO, and CSA. This publication uses UVIT data processed by the payload operations centre by the IIA. 
We thank the staff of IAO, Hanle and CREST, Hosakote, that made these observations possible. The facilities at IAO and CREST are operated by the Indian Institute of Astrophysics, Bangalore.
PKN acknowledges TIFR's postdoctoral fellowship. 
PKN also acknowledges support from the Centro de Astrofisica y Tecnologias Afines (CATA) fellowship via grant Agencia Nacional de Investigacion y Desarrollo (ANID), BASAL FB210003.

\bibliography{UV_properties_TTS}{}
\bibliographystyle{aasjournal}



\end{document}